\documentclass[10pt,letterpaper]{article}
\usepackage{opex3,amstext}
\newcommand{\la}{\langle}
\newcommand{\ra}{\rangle}
\newcommand{\xcoh}{x_\mathrm{coh}}
\newcommand{\tobj}{T_{\rm obj}}
\newcommand{\pobj}{\tilde{T}_{\rm obj}}
\newcommand{\rhoff}{\rho_{\rm SA}}

\newcommand{\lc}{l_{\rm c}}
\newcommand{\im}{i}
\newcommand{\dd}{{\rm d}}
\newcommand{\qpdc}{\delta q_{\rm PDC}}
\newcommand{\q}{{\bf q}}
\newcommand{\x}{{\bf x}}

\newcommand{\xp}{{\bf x}^{\, \prime}}

\newcommand{\sa}{{\rm SA}}
\begin{document}
\title{Ghost imaging schemes: fast and broadband}

\author{M. Bache, E. Brambilla, A. Gatti and L.A. Lugiato}
\address{INFM, Dipartimento di Fisica e Matematica, Universit{\`a}
  dell'Insubria, Via Valleggio 11, 22100 Como, Italy}
\email{morten.bache@uninsubria.it}
\homepage{http://quantim.dipscfm.uninsubria.it/quantim/}
\centerline{\small Received 29 September 2004, published 29 November 2004, Optics Express \textbf{12} (24),
  6068 (2004)}

\begin{abstract}
  In ghost imaging schemes information about an object is extracted by
  measuring the correlation between a beam that passed the object and
  a reference beam. We present a spatial averaging technique that
  substantially improves the imaging bandwidth of such schemes, which
  implies that information about high-frequency Fourier components can
  be observed in the reconstructed diffraction pattern. In the
  many-photon regime the averaging can be done in parallel and we show
  that this leads to a much faster convergence of the correlations. We
  also consider the reconstruction of the object image, and discuss
  the differences between a pixel-like detector and a bucket detector
  in the object arm. Finally, it is shown how to non-locally make
  spatial filtering of a reconstructed image. The results are
  presented using entangled beams created by parametric
  down-conversion, but they are general and can be extended also to
  the important case of using classically correlated thermal-like
  beams.
\end{abstract}

\bibliographystyle{c:/texmf/tex/latex/revtex/osa}
\bibliography{d:/Projects/Bibtex/literature}

\section{Introduction}
\label{sec:Introduction}

Lately a substantial effort has been put into understanding the physics
behind ``ghost'' imaging
\cite{klyshko:1988,belinskii:1994,strekalov:1995,pittman:1995,ribiero:1994,saleh:2000,abouraddy:2001,abouraddy:2002,gatti:2003,thermal-oe,gatti:2004a,bache:2004,ferri:2004},
%\cite{klyshko:1988,belinskii:1994,strekalov:1995,pittman:1995,ribiero:1994,saleh:2000,abouraddy:2001,abouraddy:2002,gatti:2003,gatti:2004,gatti:2004c,gatti:2004a,magatti:2004,bache:2004,bennink:2002a,bennink:2004,cheng:2004},
also called entangled imaging, two-photon imaging, coincidence
imaging, or correlated imaging. The ghost imaging schemes are based on
two correlated beams typically originating from parametric down
conversion (PDC). One beam travels a path (the test arm) containing an
unknown object, while the other beam is sent through a reference
optical system (the reference arm). Information about the object is
then obtained by measuring the spatial correlation between the beams.
This two-arm configuration allows for obtaining different kinds of
information by solely operating on the reference arm optical setup
while keeping the test arm fixed. In particular, both the image
(near-field distribution) and the diffraction pattern (far-field
distribution) of the object can be measured \cite{gatti:2003}.

The early studies concentrated on the working principles of the ghost
imaging scheme, both in terms of basic formalism
\cite{klyshko:1988,belinskii:1994} as well as the experimental
implementation \cite{strekalov:1995,pittman:1995,ribiero:1994}.  A
general theory has been developed that discusses how to extract the
information, not only in the coincidence counting regime
\cite{saleh:2000,abouraddy:2001,abouraddy:2002}, but also in the high
gain regime \cite{gatti:2003,thermal-oe,gatti:2004a,bache:2004} where
recent experimental results showed nonclassical spatial correlations
for for high-gain PDC \cite{jedrkiewicz:2004}, which promises well for
an efficient quantum ghost imaging protocol.  Most recent discussions
have instead focused on whether entanglement is necessary for
extracting the information
%\cite{abouraddy:2001,gatti:2003,gatti:2004,gatti:2004c,gatti:2004a,magatti:2004,bennink:2002a,cheng:2004},
\cite{saleh:2000,abouraddy:2001,abouraddy:2002,gatti:2003,thermal-oe,gatti:2004a,ferri:2004}
and references therein.
The present paper focuses on how to optimize any source, entangled or
not, to give as much image information as possible. The issues of
imaging bandwidth and image resolution have been taken up previously
\cite{gatti:2003,thermal-oe,gatti:2004a,bache:2004,ferri:2004}. Even
in an ideal ghost imaging system the imaging bandwidth is limited by
the source generating the correlated beams, characterized by a finite
extension of the spatial gain in Fourier space $\delta q_{\rm
  source}$: beyond this value the Fourier frequencies are cut off and
this information is lost in the reconstruction of the object
diffraction pattern. The image resolution is instead limited by the
coherence length $\xcoh$, which is given by the characteristic width
of the near-field correlation function.
%Typically, the two are connected through the relation $\xcoh=1/q_0$ \cite{thermal-oe,bache:2004}.

The question is: can we increase the source cutoff value? In the case
of PDC the cutoff is determined by the phase-matching conditions
\cite{brambilla:2004,thermal-oe,bache:2004} while in the case of a
classically correlated field generated from chaotic thermal radiation,
the bandwidth is roughly given by the inverse of speckle size found
from the self-interference of the near field \cite{goodman:1968} (see
also \cite{thermal-oe,ferri:2004}). Thus, the cutoff is an inherent
property of the source. However, in this paper we discuss a spatial
averaging technique which circumvents this cutoff and improves the
imaging bandwidth of the system, regardless of how the correlated
beams are created. The spatial average is performed over the position
of a pixel-like detector located in the test arm after the object,
exploiting that each position of this detector gives access to a
\textit{different} part of the diffraction pattern through the
correlations. Thus, making an average over all possible positions of
the test detector it is possible to substantially extend the imaging
bandwidth of the scheme.
%However, while the reconstructed
%diffraction pattern benefits from the increased bandwidth, a similar
%technique cannot be used to increase the near-field image resolution.

The spatial averaging technique works particularly well in the
high-gain regime where many photon pairs are generated in each pump
pulse. Thus the test arm has many photons per pulse transmitted by the
object, (in contrast in the low-gain coincidence counting regime only
one photon at a time is impinging on the object, and either it is
transmitted or it is not), and therefore at the measurement plane they
are scattered over the entire transverse plane. The information about
the object is then extracted by spatially correlating the intensities
recorded in the test and reference arm. In the low-gain regime this
implies registering coincidences while in the high gain regime
successive sampling over repeated shots of the pump pulse is used.
Since the spatial averaging technique employs an average over the
position of the test detector, within a single shot we can get
information about the image from all the test detector positions in
the transverse plane containing photons. This implies that in the high
gain regime a much faster convergence rate is obtained using the
spatial averaging technique.

The spatial averaging technique was already introduced in
Ref.~\cite{bache:2004} in the case where homodyne detection was used.
Since the homodyne measurements give access to orthogonal quadratures,
an increased image resolution can be obtained by using the spatial
averaging technique to measure the diffraction pattern and then use an
inverse Fourier transform to obtain the image. In the present work we
only have access to the field intensity, so we cannot use this method
to get an increased image resolution.  However, we may use a bucket
detector in the test arm when we want to observe the image, which was
not possible in the homodyne scheme due to phase-control problems, and
it turns out to make the imaging system incoherent. We discuss the
benefits and disadvantages in this respect.  We will also extend the
discussion of \cite{bache:2004} and give a more detailed and general
picture as to why the spatial averaging technique works, and as to how
much the convergence rate can be improved.

The majority of the results of this paper hold for ghost imaging
schemes in general, however we use in the following a model for PDC as
the source for the correlated beams.

The paper starts by presenting the analytical results in
Sec.~\ref{sec:Analytical-results}, and the spatial averaging scheme is
introduced and discussed. The numerical results are contained in
Sec.~\ref{sec:Numerical-examples}, going beyond the approximations
made in the analytical section and validating the results by using
very realistic parameters of current experiments. The paper is
summarized in Sec.~\ref{sec:Conclusion}.

\section{The model and analytical results}
\label{sec:Analytical-results}

We consider the PDC model for the three-wave quantum interaction
inside a $\chi^{(2)}$ nonlinear crystal previously discussed in detail
in Refs.~\cite{bache:2004,brambilla:2004}. The crystal of length $\lc$
is cut for type II phase matching, and the model consists of a set of
operator equations describing the evolution inside the crystal of the
quantum mechanical boson operators $a_j(\x,z,t)$ for the signal
($j=1$) and idler ($j=2$) fields, obeying at a given $z$ the
commutator relations $[a_i(z,\x,t),a_j^{\dag}(z,\xp,t')]
=\delta_{ij}\delta(\x-\xp)\delta(t-t')$, $i,j=1,2$. In the stationary
and plane-wave pump approximation (SPWPA) unitary input-output
transformations can be written relating the field operators in $\q$
and $\Omega$ space at the output face of the crystal
$b_j(\q,\Omega)\equiv a_j(z=\lc,\q,\Omega)$ to those at the input face
$a_j^{\mathrm{in}}(\q,\Omega)\equiv a_j(z=0,\q,\Omega)$ as follows
\begin{eqnarray}
b_{j}(\q,\Omega)=U_j(\q,\Omega)a_{j}^\mathrm{
  in} (\q,\Omega)
%\nonumber\\
+V_j(\q,\Omega)a_{k}^\mathrm{ in\dag}(-\q,-\Omega), \quad
  j\neq k=1,2.
\label{eq:inputoutput}
\end{eqnarray}
The gain functions $U_j$ and $V_j$ can for example be found in
\cite{bache:2004,brambilla:2004}.

\begin{figure}[ht] 
\centerline{    
    \scalebox{.45}{\includegraphics{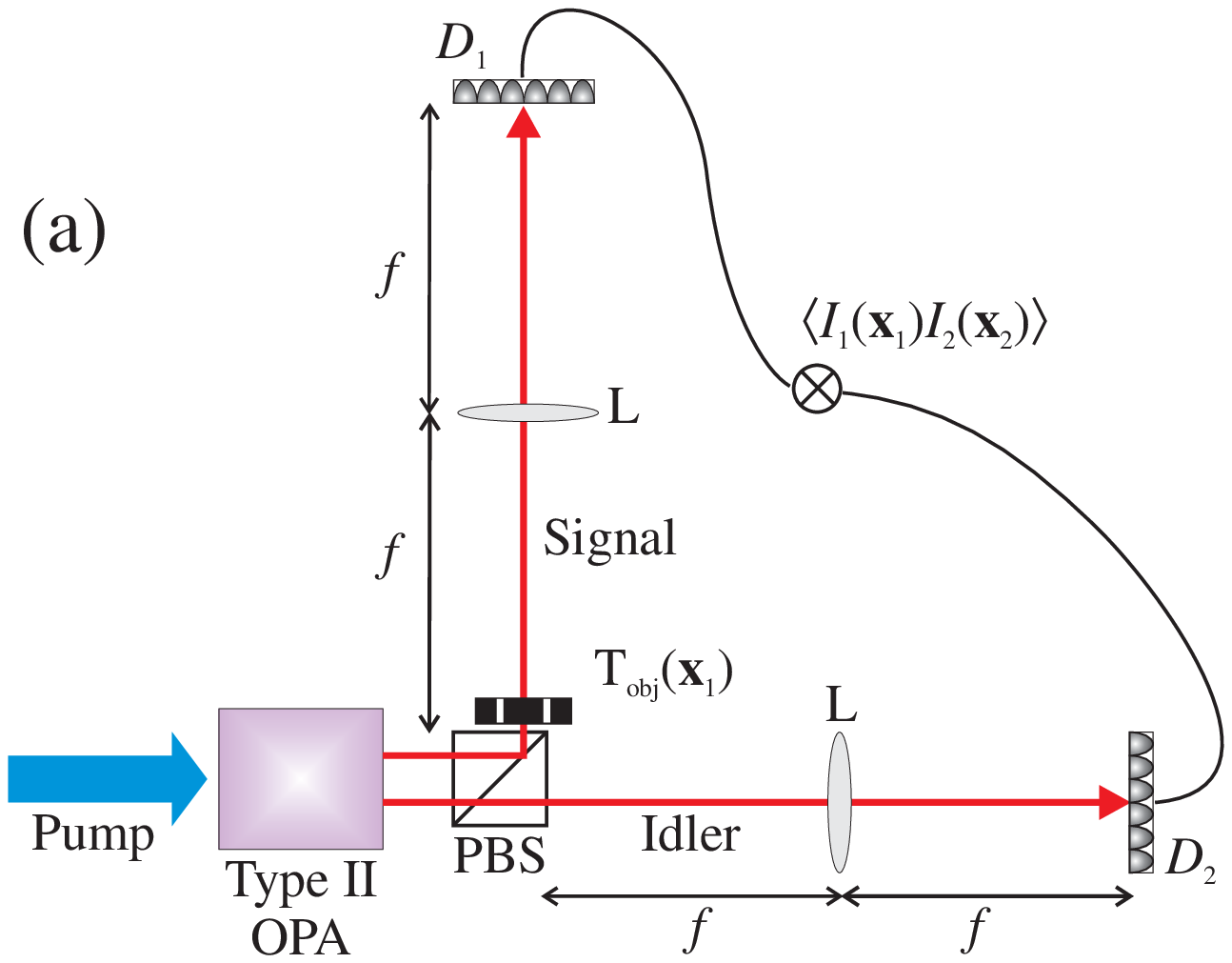}}
    \scalebox{.45}{\includegraphics{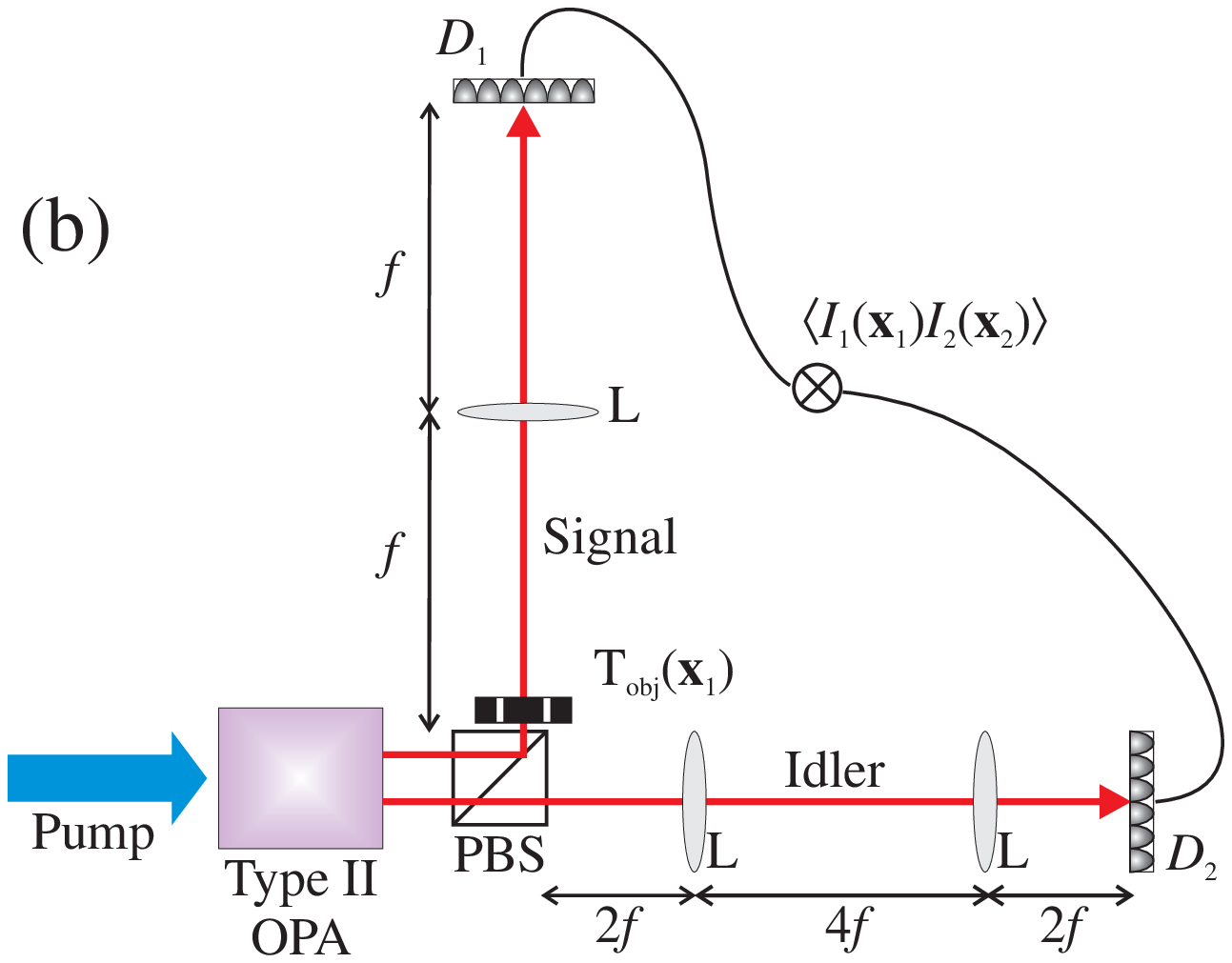}}
}
\caption{System setup. In (a) the setup for reconstructing the
  diffraction pattern of the object is shown, where the reference arm
  is in the f-f setup. In (b) the reference arm is changed to a
  telescope setup, which is used to reconstruct the object near field.
  PBS: polarizing beam splitter. L: lens of focal length $f$. $\tobj$:
  object. $D_1$, $D_2$: arrays of detectors.}
\label{fig:setup}
\end{figure}  

The system setup we consider is the same as in Refs.
\cite{gatti:2003,thermal-oe,gatti:2004a,bache:2004} and has the
characteristic two-arm configuration of the ghost imaging scheme, see
Fig.~\ref{fig:setup}.  The signal and idler exiting from the crystal
are separated with a polarizing beam splitter (PBS). The object is
located in the test (signal) arm immediately after the PBS, and the
transmitted field then propagates through an f-f lens system: a lens
placed distance $f$ from the object as well as to the measurement
plane, where $D_1$ (an array of pixels or a CCD camera) is recording
the intensity of the field. Thus $D_1$ is in the focal plane of the
lens, but it is important to stress that it cannot on its own measure
the interference pattern of the object because the signal beam itself
is incoherent. This test arm setup is kept fixed, while in the
reference (idler) arm two different configurations are used.  In
Fig.~\ref{fig:setup}(a) the reference arm contains also an f-f lens
system, and by recording the idler field intensity with $D_2$ (another
CCD camera) and correlating it with the signal intensity, information
about the object diffraction pattern can be extracted from the
correlations.  Conversely, in Fig.~\ref{fig:setup}(b) a so-called
telescope setup is used consisting of two lenses with focal length $f$
placed 4$f$ from each other and $2f$ from the PBS and $D_2$,
respectively. This setup allows for retrieval of the object image
distribution from the correlations.

The analytical results for the correlations in the two cases of
Fig.~\ref{fig:setup} were derived in
Ref.~\cite{gatti:2003,thermal-oe}, and we review them in the
following. For simplicity we neglect for the moment the temporal
argument, which corresponds to using a narrow frequency filter after
the crystal. The propagation of the fields from the PBS to the
measurement planes are described by the kernels $h_j(\x_j,\xp_j)$. The
fields at the measurement plane $c_j(\x_j)$ are then found by the
Fresnel transformation $c_j (\x_j) = \int \dd \xp_j h_j(
\x_j,\xp_j) b_j (\xp_j) + L_j (\x_j)$. $L_j(\x_j)$
represent the losses in the propagation which are proportional to the
vacuum field operators and are uncorrelated from $b_j(\x_j)$. The
field intensities at the measurement planes
$I_j(\x_j)=c_j^\dagger(\x_j)c_j(\x_j)$ are detected with $D_j$, and
result in the correlation $\la I_1(\x_1)I_2(\x_2)\ra=\la
c_1^\dagger(\x_1)c_1(\x_1) c_2^\dagger(\x_2)c_2(\x_2) \ra$. The
information about the object is obtained by subtracting the background
term to obtain the intensity fluctuation correlations
\begin{equation}
G(\x_1, \x_2) = \langle I_1 (\x_1) I_2 (\x_2) \rangle - \langle I_1
(\x_1)\rangle \langle I_2 (\x_2) \rangle.
\label{eq:G}
\end{equation}
It is then straightforward to show the following essential result
\cite{gatti:2003,thermal-oe} 
\begin{equation}
G(\x_1, \x_2)  = \left| \int \dd \xp_1
\int \dd \xp_2  h_1 (\x_1, \xp_1) h_2 (\x_2, \xp_2) 
%\langle b_1 (\xp_1) b_2(\xp_2) \rangle
\Gamma(\xp_1-\xp_2,\Omega=0)
\right|^2,
\label{eq:Gfinal}
\end{equation}
Here $\Gamma(\x_1,\x_2,\Omega)$ is the near-field correlation at the
crystal exit, and in the SPWPA and with $a_j^{\mathrm{in}}$ in the
vacuum state it is found from Eq.~(\ref{eq:inputoutput}) as
\begin{eqnarray}
\Gamma(\x_1,\x_2,\Omega)\equiv \int \dd \tau
e^{-i\Omega \tau} 
\langle b_1 (\x_1,t_1) b_2(\x_2,t_1+\tau) \rangle 
%\nonumber\\
=
\int \frac{ \dd \q } {(2\pi)^2} e^{\im \q \cdot (\x_1-\x_2)}
\gamma(\q,\Omega)
\: ,
\label{eq:gammaentangled}
\end{eqnarray}
where we have introduced the gain function $\gamma(\q,\Omega)=U_1
(\q,\Omega) V_2 (-\q,-\Omega)$. Since Eq.~(\ref{eq:gammaentangled}) is
a function of $\x_1-\x_2$ (because of the SPWPA) we will use the
notation $\Gamma(\x_1-\x_2,\Omega)=\Gamma(\x_1,\x_2,\Omega)$ in the
following.

We should mention that when temporal argument is taken into account,
Eq.~(\ref{eq:Gfinal}) is no longer valid. We consider the intensities
averaged over the detection time $T_{\rm D}$ as $I_j(\x_j)=T_{\rm
  D}^{-1}\int_{T_{\rm D}} \dd t c_j^\dagger (\x_j,t) c_j(\x_j,t)$.
When $T_{\rm D}$ is much larger than the coherence time of the source
$\tau_{\rm coh}$ (which for PDC is typically the case) the following
expression holds instead \cite{gatti:2004a}
\begin{equation}
G(\x_1, \x_2)  =\frac{1}{T_{\rm D}}\int \frac {\dd \Omega}{2\pi} \left|
 \int \dd \xp_1
\int \dd \xp_2  h_1 (\x_1, \xp_1) h_2 (\x_2, \xp_2) 
%\langle b_1 (\xp_1) b_2(\xp_2) \rangle
\Gamma(\xp_1-\xp_2,\Omega)\right|^2.
\label{eq:Gtime}
\end{equation}
This form will be used later for comparing with the numerics.

We fix the test arm once and for all as shown in Fig.~\ref{fig:setup},
so $h_1(\x_1,\xp_1)\propto e^{-\im \x_1 \cdot \xp_1 k/f }
\tobj(\xp_1)$, where $k$ is the free space wave number of the
degenerate down-converted fields.  To extract information about the
diffraction pattern the reference arm is set in
the f-f setup of Fig.~\ref{fig:setup}(a), $h_2(\x_2,\xp_2)\propto e^{-
  \im \x_2 \cdot \xp_2 k/f}$, and Eq.~(\ref{eq:Gfinal}) then
straightforwardly implies the correlation
\cite{gatti:2003,thermal-oe,gatti:2004a}
\begin{equation}
G_{\rm{f}}(\x_1, \x_2) 
%=\frac{(2\pi)^2TR}{(\lambda f)^4}
\propto \left| \gamma( -\x_2 k/f,\Omega=0 )
% V_2 ( \x_2  k/f )   
\,  
\pobj \left[ (\x_1 +\x_2)k/ f  \right]\right|^2 \: ,
\label{eq:diffpatt_ent}
\end{equation}
where $\pobj (\q) = \int \frac {\dd \x}{2\pi} e^{-\im \q \cdot \x}
\tobj(\x)$ is the amplitude of the object diffraction pattern. The
subscript ``f'' denotes that the reference arm is in the f-f
configuration. The correlation provides information about the
diffraction pattern of the object if we fix $\x_1$ and scan $\x_2$,
but since the gain also depends on $\x_2$ there is a limit to the
information we can extract.  Precisely, the gain $\gamma( -\x_2
k/f,\Omega=0 )$ introduces a cutoff of the reproduced spatial Fourier
frequencies at a certain characteristic value which we denote $\delta
q_{\rm PDC}$; the \textit{imaging bandwidth} of the PDC source.

We will now show how to circumvent this source-related limitation on
the imaging bandwidth. As previously shown \cite{bache:2004} we may in
a suitable way average over the position of the test arm detector
$\x_1$: 
%, and from the structure of Eq.~(\ref{eq:diffpatt_ent}) we should
%do it as follows. 
First, a change of variables is introduced as $\x\equiv \x_1+\x_2$,
and then an average over $\x_1$ is performed. We then obtain
\begin{eqnarray}
  \label{eq:sa-conv}
  G_{\mathrm{f},\sa}(\x)\equiv\int \dd \x_1
  G_{\mathrm{f}}(\x_1,\x-\x_1)\propto \int \dd \x_1
\left| \gamma[ (\x_1-\x) k/f,\Omega=0 ]\,
\pobj (\x k/ f  )\right|^2
\nonumber
\\
= |\pobj (\x k/ f  )|^2\int \dd \x_1
\left| \gamma[ (\x_1-\x) k/f,\Omega=0 ]\right|^2
\simeq{\rm const}\times|\pobj (\x k/ f
  )|^2
%\int \dd \x_1[\langle I_1 (\x_1) I_2 (\x-\x_1) \rangle -
%  \langle I_1 (\x_1)\rangle \langle I_2 (\x-\x_1) \rangle]
  \label{eq:sa}
\end{eqnarray}
where the subscript ``SA'' indicates that a spatial average has been
carried out. The final approximation in~(\ref{eq:sa}) is that
$|\gamma(\q,\Omega=0)|^2$ is a bound function inside the detection
range of $\x_1$ implying that the integral evaluates to a constant.
Thus, there is now no gain cutoff of the diffraction pattern, so the
imaging bandwidth becomes practically infinite (apart from limitations
arising from the finite size of the optical elements).\footnote{We
  should mention that when $\x$ is taken to the boundary of the
  detection range, the integral is no longer a constant. Thus at the
  boundaries the bandwidth slowly dies out, but it is a quite small
  effect only affecting a range of $\delta q_{\rm PDC}$ there. In our
  numerical simulations we use periodic boundary conditions, so this
  limit does not come into play.} Note that this average over $\x_1$
does not correspond to $D_1$ being a bucket detector. Instead, the
change in variables $\x\equiv \x_1+\x_2$ implies that the spatial
averaging technique is a convolution of the signal and idler intensity
fluctuations, which
in practice is easily calculated using fast Fourier transform
techniques.

\begin{figure}[ht] 
\centerline{    
    \scalebox{.55}{\includegraphics{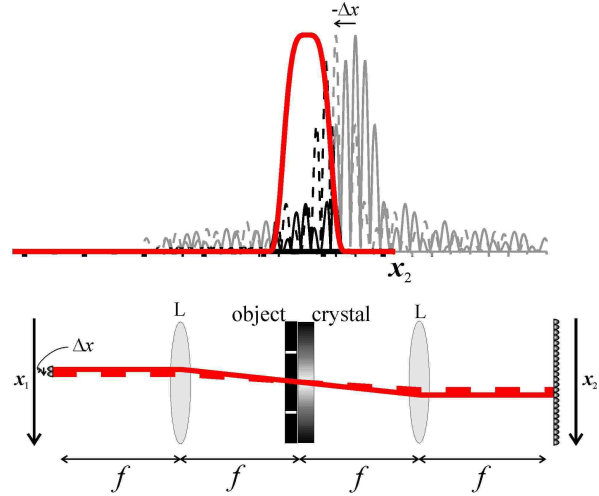}}
}
\caption{The spatial averaging technique explained by the Klyshko
  picture. The lower sketch shows the unfolded version of
  Fig.~\ref{fig:setup}(a). The red line represents the ray from the
  detector at position $x_1^0$ as it travels through the system and
  reaches its conjugate point in the reference plane on the right
  side. The upper plot shows the source gain curve in Fourier space
  and the position of a typical diffraction pattern for the current
  value of $x_1^0$; only the black part is amplified while the gray
  part is not amplified. As the test detector position is changed to
  $x_1^0+\Delta x$ the dashed red ray comes into play, and the
  diffraction pattern in the upper plot moves a corresponding amount
  (also dashed). The gain instead remains the same, and hence a
  different part of the diffraction pattern is amplified. (Movie size
  449 KB).}
\label{fig:klyshko}
\end{figure}  

In Fig.~\ref{fig:klyshko} we give an intuitive explanation of why the
bandwidth is extended by averaging over $\x_1$, and it is based on the
Klyshko picture \cite{klyshko:1988} (see also \cite{tan:2003}). This
states that the propagation in the two distinct arms can be viewed in
an ``unfolded'' scheme having the nonlinear crystal as the hinge point
of the signal and idler rays, i.e. the crystal can be treated as a
geometrical reflection mirror. This is a consequence of momentum
conservation in the PDC process, so a signal ray with momentum $\q'$
implies an idler ray with momentum $-\q'$. Thus, we may treat the
scheme starting from the signal detector with the signal ray passing
backwards in time through the test arm, until it encounters the
crystal and becomes the idler ray traveling forward in time. The point
is that the action of the crystal is to fix the idler ray propagation
direction through the momentum conservation, and in the unfolded
version of Fig.~\ref{fig:klyshko} the idler ray is a straight-line
continuation of the signal ray.
% Effectively, we view the system starting from the
% test detector and propagating towards the reference detector. In our
% case, the ray propagates from the test detector through the test arm
% (backwards in time) encounters the object, then the crystal, and the
% ray becomes the idler propagating through the reference arm to the
% detector (forward in time). 
According to the Klyshko picture the detector in the test arm acts as
a point-like source emitting a spherical wave from $\x_1$. This wave
is transformed by the f-f lens system into a tilted plane-wave with
transverse wave vector $\q=-\x_1 k/f$ which then after impinging on
the object is diffracted. Inside the crystal the diffracted signal ray
is ``converted'' into the idler ray with ``transmission'' function
$\gamma(\q,\Omega=0)$ determined by the phase-matching conditions.
This transmission function $\gamma(\q,\Omega=0)$ is more precisely
amplifying certain components of the signal ray and damping others
(gray area in Fig.~\ref{fig:klyshko}), and this is the origin of the
infamous cutoff.  Therefore, as the idler ray now propagates forward
through the f-f lens system in the reference arm we can with $D_2$
only detect a bandwidth-limited version of the diffraction pattern
$\left| \gamma( -\x_2 k/f,\Omega=0 ) \pobj \left[ (\x_2 +\x_1)k/ f
  \right]\right|^2$.
% With the f-f setup in the test arm the transmitted signal field is in
% the far field, which means that fixing a certain test detector
% position $\x_1^0$ means fixing a certain wavenumber. The backward
% propagation through the f-f lens system makes this a tilted plane wave
% at the object, after which the crystal is encountered that
% ``converts'' the signal to the idler ray, but with a certain gain
% bandwidth that acts as the infamous cutoff. This idler ray carries
% then a bandwidth-limited version of the Fourier components of the
% transmitted field from the object, and through the f-f lens system in
% the reference arm the detection is in the focal plane: we detect the
% bandwidth-limited Fourier components of the diffraction pattern.  
The spatial averaging technique exploits that as we move to a new
detector position $\x_1^0+\Delta \x$ the center of the object changes
correspondingly to $\x_2=-\x_1^0-\Delta \x$, while the gain remains
fixed [due to the structure of Eq.~(\ref{eq:diffpatt_ent})]. Thus, by
changing the test detector position, we measure a different part of
the spatial Fourier spectrum, as shown in Fig.~\ref{fig:klyshko}.
Therefore, by performing a suitable average as described in
Eq.~(\ref{eq:sa}) we cover the entire Fourier plane, and effectively
the bandwidth has become unlimited.
% Since this technique averages not only over independent shots but also
% over detector positions, a proviso for this to work is that two
% neighboring detectors are statistically independent. In other words,
% the characteristic autocorrelation length in the far-field should be
% on the order of the pixel size, which effectively sets a limit to the
% pump beam waist. If this is not the case the statistical averages are
% not independent and the convergence rate is reduced.

Additionally, the spatial averaging technique gives an increased
convergence rate of the correlation. This is again related to the fact
that when the test detector position is changed from $\x_1^0$ to
$\x_1^0+\Delta \x$ the gain does not change position but the
diffraction pattern does.  Assuming that the PDC gain extension is much larger
than the extension of a pixel, the shifted diffraction
pattern at $\x_1^0+\Delta \x$ overlaps quite substantially with the
previous one. Thus, as $\x_1$ is scanned a given position of the
diffraction pattern has as many independent contributions as there are
independent modes inside the gain bandwidth, which as a good estimate
is given by the ratio of the PDC bandwidth $\qpdc$ with
the pump bandwidth $\delta q_{\rm p}=2/w_0$ \cite{brambilla:2004},
where $w_0$ is the pump waist. A measure of the speedup in the
correlation convergence in each transverse dimension is therefore
given by
\begin{equation}
  \label{eq:n_conv}
  \rhoff%_{\rm conv}^{\rm FF}
=\qpdc/\delta q_{\rm p}.
\end{equation}
In the simulations shown later concerning the convergence rate, we
used a temporal grid that corresponds to a 22 nm interference filter.
In that case $\qpdc\simeq 6 q_0$, where $q_0=\sqrt{k/\lc}$ is the
natural bandwidth of PDC at a given temporal frequency
\cite{brambilla:2004}. The pump size in the numerics was chosen so
$\delta q_{\rm p}=q_0/18$ (corresponding to a pump size of
600~$\mu$m), implying we should expect a convergence rate speedup of
$\rhoff\simeq 10^2$.
% When considering only degenerate contributions (i.e. neglecting
% temporal contributions as we have done so far in the analytical
% section) $\qpdc \simeq q_0$ where $q_0$ is the natural bandwidth of
% PDC at a given temporal frequency, and it is roughly given by
% $q_0^2=k/\lc$ \cite{brambilla:2004}. In the simulations with time a
% temporal grid was used that corresponds to a 22 nm interference
% filter. In this case $\qpdc\simeq 6 q_0$ and since we used $\delta
% q_{\rm p}=q_0/18$ (corresponding to a pump size of 600~$\mu$m) we get
% $\rhoff\simeq 10^2$.

% Another advantage
% of this spatial averaging technique is that at $\x_1^0+\Delta \x$ the
% shift of the diffraction pattern position amplified by the gain
% overlaps quite substantially with the previous region (since the gain
% extension $q_0$ is much larger than the extension of a pixel $\Delta
% \x$). If the neghboring pixels are independent, then we effectively at
% the new position $\x_1^0+\Delta \x$ get new information about the
% overlapping region that is statistically independent from the previous
% detector position.  Eventually this means that the convergence of the
% correlation is sped up by a factor $n^D$, where $n$ is the number of
% independent pixels in the dimension, and $D$ is the number of
% transverse dimensions involved in the problem. This can be a quite
% substantial increase in the convergence rate, especially when we
% consider 2 transverse dimensions.
 
Note that instead of fixing $\x_1$ and scanning $\x_2$, we may scan
$\x_1$ and fix $\x_2$. In this case Eq.~(\ref{eq:diffpatt_ent})
reveals that the gain no longer limits the imaging bandwidth, and
therefore it is in principle not necessary to do a spatial average to
overcome the gain limitation. The physical explanation behind this is
again found from the Klyshko picture: fixing $\x_2$ at a suitable
position (i.e. at the position of maximum gain, cf.
Fig.~\ref{fig:klyshko}), scanning $\x_1$ will move the diffraction
pattern seen at this position over the whole
range giving an unlimited imaging bandwidth. % Starting
% from the fixed reference detector position $\x_2^0$ it is in the focal
% plane, and thus selects a given wave number, which propagates
% (backwards in time) through the f-f lens system in the reference arm
% to become a tilted plane wave. Now it encounters the crystal which
% turns the idler ray into the signal ray with a certain gain bandwidth,
% but in contrast to before the beam now consists only of the single
% wave number selected by the position $\x_2^0$. If this wave
% number is inside the gain, the plane wave is merely amplified and
% impinges on the object. But in contrast to before, now there is
% nothing after the object that limits the bandwidth, and the
% transmitted field from the object may therefore undisturbed interfere
% through the f-f lens system.  Ultimately, we may measure by scanning
% $\x_1$ a diffraction pattern with an unlimited bandwidth. Thus,
% the difference between the two cases lies in the position of the
% object with respect to the crystal. 
However, scanning $\x_1$ and fixing $\x_2$ does not benefit of the
improved convergence. To achieve this a spatial average should be
done, whereby the method amounts to the same operation as described
previously.

Let us now turn to the problem of reconstructing the object image.
Keeping the test arm fixed and changing the reference arm to the
telescope setup [see Fig.~\ref{fig:setup}(b)], $h_2(\x_2,\xp_2)=
\delta(\x_2-\xp_2)$ and Eq.~(\ref{eq:Gfinal})
becomes \cite{gatti:2003,thermal-oe,gatti:2004a}
\begin{eqnarray}
G_{\rm T}(\x_1,\x_2)
\propto
\left| \int \dd \xp_1
% \la b_1(\xp_1)b_2(\x_2)\ra
\Gamma(\xp_1-\x_2,\Omega=0)
 \tobj \left( \xp_1  \right) e^{-\im \xp_1 \cdot \x_1 k/f}
\right|^2
\label{eq:telescope1} \\
\approx
|\tobj(\x_2)|^2 \left|\gamma \left (\x_1 k/f,\Omega=0\right)\right|^2 .
\label{eq:telescope}
\end{eqnarray}
Thus, fixing $\x_1$ the object can be observed by scanning $\x_2$. The
approximation that leads to Eq.~(\ref{eq:telescope}) holds as long as
the smallest length scale of the object is larger than the coherence
length $\xcoh$ of $\Gamma(\xp_1-\x_2,\Omega=0)$. This is because
$\Gamma(\xp_1-\x_2,\Omega=0)$ is nonzero in a region of the size
$\xcoh$ around $\xp_1=\x_2$ and thus $\tobj$ changes slowly with
respect to this function and may consequently be pulled out of the
integration. In general, Eq.~(\ref{eq:telescope1}) is a convolution
between the correlation function and the object, and hence the width
$\xcoh$ of the near-field correlation function
$\Gamma(\xp_1-\x_2,\Omega=0)$ determines the resolution of the
reconstructed image, and has typically a value of $\xcoh=1/q_0$.

In reconstructing the image a technique corresponding to the spatial
average done for reconstructing the diffraction pattern would result
in a largely increased image resolution. Unfortunately, it is not
possible to carry out such a spatial average to achieve this. However,
if we make a simple sum over all the positions of $D_1$ (corresponding
to $D_1$ being a bucket detector), instead of
Eq.~(\ref{eq:telescope1}) we have the following exact
expression\footnote{The expression is exact because no approximations
  have been made about object length scales vs. $\xcoh$.}
\begin{equation}
\bar G_{{\rm T}}(\x_2)  = \int \dd \x_1 G_{\rm T}(\x_1,\x_2)\propto
\int \dd \x_1 |
% \la b_1(-\x_1)b_2(-\x_2)\ra
\Gamma(\x_1-\x_2,\Omega=0)
|^2|
 \tobj \left( \x_1  \right) |^2,
\label{eq:telescope1-bucket}
\end{equation}
where the bar denotes that $D_1$ is a bucket detector and therefore
that $\x_1$ has been integrated out. Eq.~(\ref{eq:telescope1-bucket})
has the form of an incoherent imaging scheme, while
Eq.~(\ref{eq:telescope1}) has the form of a coherent imaging scheme
(the same conclusion -- that using a bucket detector can make the imaging
incoherent -- was reached in Ref.~\cite{abouraddy:2002} in the
low-gain regime). As we shall see later, the incoherent form has both 
advantages and drawbacks compared to the coherent case.

\section{Numerical examples}
\label{sec:Numerical-examples}

The imaging performance of the system was checked through numerical
simulations of the model described in detail in Ref.
\cite{brambilla:2004}. It includes spatial and temporal dispersion up
to second order, as well as a Gaussian shape in space and time of
the pump beam.  The simulations with 1 transverse dimension (1D) were
calculated including the temporal argument (using a grid of $N_x=512$
and $N_t=64$), while the ones with 2 transverse dimensions (2D) had a
more qualitative nature since they neglected the time argument (a grid
of $N_x=N_y=256$ was used).\footnote{This was done to save CPU time
  since many thousands of repeated pump shots were needed for the
  correlations to converge, and corresponds to using a narrow
  bandwidth interference filter after the crystal. In fact, the
  temporal grid acts just as an interference filter by providing a
  cutoff in frequency space. In the simulations with time the cutoff
  was chosen so it corresponds to having a 22 nm interference filter
  after the BBO crystal, while the simulations neglecting time are
  equivalent to placing a $<1$ nm
  interference filter after the crystal. % The approach was justified by
%   comparing quantitatively simulations with and without time for a
%   fewer number of pump shots.
} Each pump shot was propagated along the crystal in $N_z=200$ steps.
The Gaussian pump profile had a waist $w_0=600~\mu$m and a duration of
$\tau_0=1.5$~ps, and the other parameters were as in
\cite{bache:2004,brambilla:2004} chosen to closely mimic that of an
$\lc= 4$~mm long BBO crystal used in a current experiment in Como
\cite{jedrkiewicz:2004}. The characteristic space and time units of
the PDC source are $x_{\rm coh}=1/q_0\simeq 16~\mu$m and $\tau_{\rm
  coh}=1/\Omega_0\simeq 0.96$~ps. The pulsed character of the pump is
important because typically the CCD cameras used in experiments have
very slow response times. If the measurement time becomes too long
with respect to $\tau_{\rm coh}$, the visibility of the correlation
becomes very poor (see also
Refs.~\cite{saleh:2000,thermal-oe,gatti:2004a}). This only affects the
case when the intensity is detected in the high gain regime, i.e. when
the background term in Eq.~(\ref{eq:G}) is substantial, while it does
not affect the coincidence counting regime or the case when homodyne
measurements are performed as in Ref.~\cite{bache:2004}.  Note that in
the telescope setup the performance was optimized by taking the
imaging plane of the telescope setup inside the crystal by the amount
$\Delta z=-(1/n_1+1/n_2)\tanh(\sigma_{\rm p} \lc)/\sigma_{\rm p}$
\cite{bache:2004,brambilla:2004}, where $n_j$ are the refractive
indices, and $\sigma_{\rm p}$ is a gain parameter. In this way, the
quadratic phase dependence of the gain is cancelled. For more
technical details on this and the numerics we refer to
Refs.~\cite{bache:2004,brambilla:2004}.

\begin{figure}[ht]
\centerline{    
    \scalebox{.55}{\includegraphics{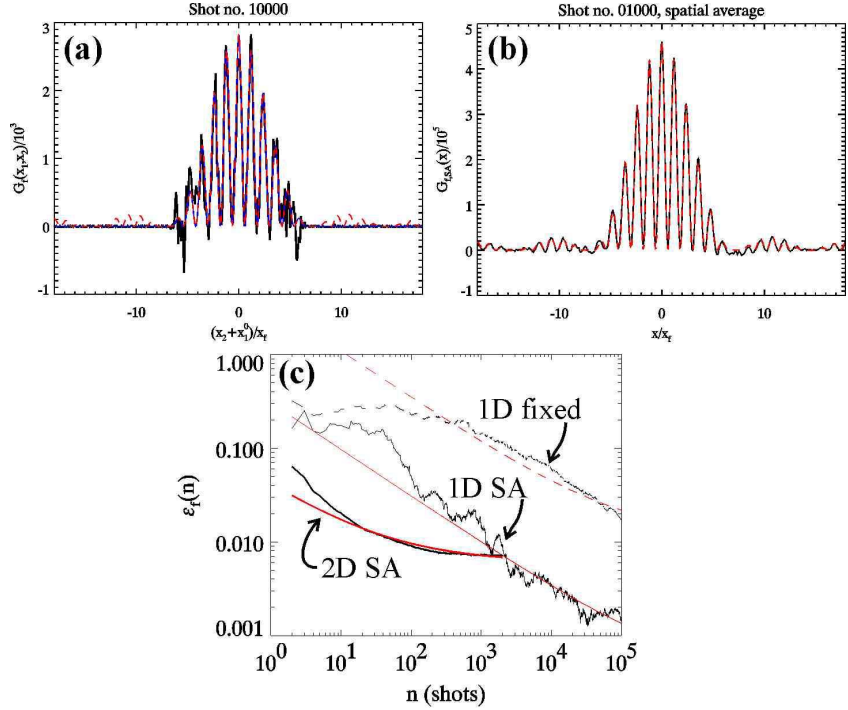}}
%     \scalebox{.29}{\includegraphics{../../OPA/GhostImage/Simulations/1D03/final-ff.ps}}
%     \scalebox{.29}{\includegraphics{../../OPA/GhostImage/Simulations/1D03a/final-ff-sa.ps}}
%     \scalebox{.54}{\includegraphics{conv1D2Dff.eps}}
}
\caption{Reconstructing the diffraction pattern using the f-f setup in the
  reference arm. (a) shows the correlation from a fixed $x_1$ after
  averaging over 10000 shots while (b) shows the correlation using the
  spatial averaging technique and averaging over 1000 shots. In (a) and
  (b) black curves are numerics, while dashed red curves are $\pobj$.
  In (a) the blue curve is the analytically calculated correlation in
  the SPWPA in arbitrary units [Eq. (\ref{eq:diffpatt_ent}) including
  integration over time, as shown in Eq.~(\ref{eq:Gtime})]. (c) shows
  $\varepsilon(n)$ from Eq.~(\ref{eq:1}) in a double-logarithmic plot
  using the spatial averaging technique (full) and a fixed detector
  (dashed), while the red curves are fits to Eq.~(\ref{eq:exp}).
  Double slit parameters: aperture 14 $\mu$m (5 pixels) and aperture
  center distance 87 $\mu$m (30 pixels). [Movie sizes: (a) 1 MB and
  (b) 933 KB].}
\label{fig:ds}
\end{figure}  

Let us show quantitatively how the object diffraction pattern is
reconstructed by using a double slit as an object in the setup of
Fig.~\ref{fig:setup}(a). The correlations as calculated in 1D both
with and without the spatial average are presented in
Fig.~\ref{fig:ds}(a)-(b). The correlation without spatial average in
(a) clearly suffers from a limited bandwidth, and from the animation a
slow convergence rate is evident. In contrast, the correlation with
spatial average in (b) is able to reproduce the entire spectrum of the
diffraction pattern and after much fewer repeated pump shots. To check
the convergence rate, we have calculated the root-mean-squared error
of the correlation $G_{\rm {f}}^{n}$ (after averaging over $n$ number
of shots) relative to the analytically calculated correlation $G_{\rm
  {f}}$.  $G_{\rm {f}}$ was calculated by using semi-analytical
methods including the analytical SPWPA gain as well as integrating
over time [see Eq.~(\ref{eq:Gtime})]. The error is then evaluated as
\begin{equation}
  \label{eq:1}
\varepsilon_{\rm f}(n)=
\left(\Sigma_{\x}  |G_{\rm {f}}^{n}(\x)-G_{\rm
  {f}}(\x)|^2\right)^{1/2}
\end{equation}
where at each shot the maximum of $G_{\rm {f}}^{n}$ is rescaled to the
maximum of $G_{\rm {f}}$. Figure~\ref{fig:ds}(c) shows $\varepsilon
(n)$ and clearly the fixed detector case (dashed) converges much
slower than the spatial average (full).
% Moreover the spatial average ends up at a much lower level, i.e. a
% smaller error, which is a consequence of the increased bandwidth. 
The red curves are fits to the function
\begin{equation}
  \label{eq:exp}  
  \varepsilon_{\rm fit}(n)=(d_0 n)^{-1/2}+d_1
\end{equation}
with $n$ being the number of shots. The factor $d_0$ has the dimension
shot$^{-1}$ and the ratio between $d_0$'s obtained using the spatial
average and using fixed detector should then give an indication of the
increased convergence rate. From the fits of Fig.~\ref{fig:ds}(c),
$d_0^{\sa,\rm 1D}/d_0^{\rm fixed,\rm 1D}\simeq 125$, which corresponds
well to the value $\rhoff\simeq 10^2$ estimated in the previous
section. The factor $d_1$ indicates the error remaining after
convergence, but in this case the correlations are not converged
sufficiently within the number of shots shown.  Now, according to what
was predicted in Eq.~(\ref{eq:n_conv}), using the spatial average in
2D should benefit with a factor of $\rhoff$ in convergence speedup
compared to using the spatial average in the 1D case (due to the extra
dimension to average over). In Fig.~\ref{fig:ds}(c) the result of such
a 2D simulation is also shown,\footnote{This was a full 3+1D simulation
  with $N_x=256$, $N_y=128$, $N_z=200$ and $N_t=32$.  The slits had
  the same extension in the $x$-direction as in the 1D case and were
  580 $\mu$m (90 pixels) long in the $y$ direction.} and the improved
convergence rate of the 2D simulation is evident: from the fits
$d_0^{\sa,\rm 2D}/d_0^{\sa,\rm 1D}\simeq 73$, again corresponding well
to the predicted estimate of $\rhoff\simeq 10^2$. A minor problem in
this comparison is that the 2D results saturate very quickly to a
rather high error level of 1\%, while the 1D results go as low as
0.1\%.  This difference turned out to be caused by the Gaussian shape
of the pump field and the object extension; the object is very
localized in the $x$ direction and therefore we get a low error rate
in the 1D case. However, in the 2D case the object is quite extended
in the $y$ direction and the average error reported in the
$\varepsilon$ value is therefore higher. This quick saturation in 2D
gave some numerical problems in fitting well the data to
Eq.~(\ref{eq:exp}), so the $d_0$ value obtained should be taken
cautiously. However, there is no doubt about the main point: going
from 1D to 2D the spatial averaging technique speeds further up.
% The advantage of this comparison is obviously that
% they converge to the same state, which can be appreciated in the right
% plot of Fig.~\ref{fig:ds} where we have included the results of such a
% 2D simulation.\footnote{This was a full 3+1D simulation with
%   $N_x=256$, $N_y=128$, $N_z=200$ and $N_t=32$.  The slits had the
%   same extension in the $x$-direction as in the 1D case and were 580
%   $\mu$m long (90 pixels).} The improved convergence rate of the 2D
% simulation is evident, and the $d_1^2$ factor of the fit is 17 times
% larger than the 1D case which again agrees very well with the
% predicted $\rhoff$. Finally, it is evident by eye that the convergence
% point (i.e. where the convergence curve becomes flat) between the
% three cases have roughly a factor of $\rhoff$ between them. This was
% confirmed by a more detailed analysis of the fits.

\begin{figure}[ht] 
\centerline{    
    \scalebox{.65}{\includegraphics{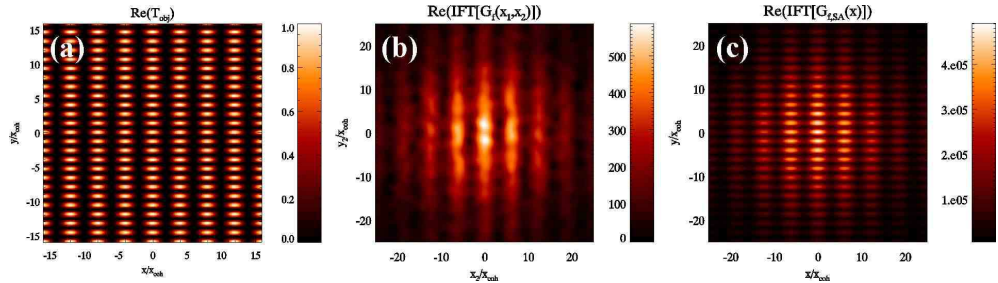}}
%     \scalebox{.29}{\includegraphics{../../OPA/GhostImage/Simulations/2D22e/pattern.ps}}
%     \scalebox{.29}{\includegraphics{../../OPA/GhostImage/Simulations/2D22e/20000-36000/ift-fix.ps}}
%     \scalebox{.29}{\includegraphics{../../OPA/GhostImage/Simulations/2D22e/20000-36000/ift-sa.ps}}
%%     \scalebox{.29}{\includegraphics{../../OPA/GhostImage/Simulations/2D22b/pattern.ps}}
%%     \scalebox{.29}{\includegraphics{../../OPA/GhostImage/Simulations/2D22b/ift-fix.ps}}
%%     \scalebox{.29}{\includegraphics{../../OPA/GhostImage/Simulations/2D22b/ift-sa.ps}}
}
\caption{Reconstructing the diffraction pattern of the object shown in
  (a) which has spatial Fourier components both inside and outside the
  PDC bandwidth. The real part of the inverse Fourier transform of the
  reconstructed diffraction pattern using the f-f setup is shown with
  (b) fixed $\x_1$ and (c) using the spatial averaging technique.
  $35000$ shots were used. }
\label{fig:cosine}
\end{figure}  

Let us show an example where the reconstructed diffraction pattern
using the f-f setup leads to completely different results depending on
whether the spatial average is applied or not. The object is shown in
Fig.~\ref{fig:cosine}(a) and is a 2D mask $T_{\rm
  obj}(\x)=[1+\cos(xq_0)][1+\cos(3yq_0)]/4$ (consisting of 4 main
sidebands, 2 located at $\q/q_0=\pm {\bf e}_x$ and 2 located at
$\q/q_0=\pm 3{\bf e}_y$).  We have chosen the peaks in the
$y$-direction so they lie outside the imaging bandwidth of the source,
while the peaks in the $x$-direction lie inside the bandwidth. % Hence
% the correlations without spatial average should reproduce what
% corresponds to a roll-pattern, while only the true square pattern is
% revealed by the correlations with the spatial average. This picture is
% confirmed by 
In the numerics the diffraction pattern was reconstructed from the
correlation with $\x_1$ fixed as well as employing the spatial
averaging technique. Instead of showing these data, we have in
Fig.~\ref{fig:cosine}(b) and (c) shown the results after taking the
inverse Fourier transform (IFT) of the reconstructed diffraction
patterns. In this way we can reconstruct to some extend the nature of
the near field mask (some phase information is lost, obviously, but in
this simple case the lost phase information is not so important).
Figure~\ref{fig:cosine}(b) shows ${\rm Re}({\rm IFT}[G_{\rm
  f}(\x_1,\x_2)])$, i.e. without spatial average, and only the roll
pattern in the $x$-direction is reproduced.  Instead, the correct
object is reproduced by the result ${\rm Re}({\rm IFT}[G_{{\rm
    f},\sa}(\x)])$ from using the spatial average, as shown in
Fig.~\ref{fig:cosine}(c).

\begin{figure}[ht] 
\centerline{    
    \scalebox{.65}{\includegraphics{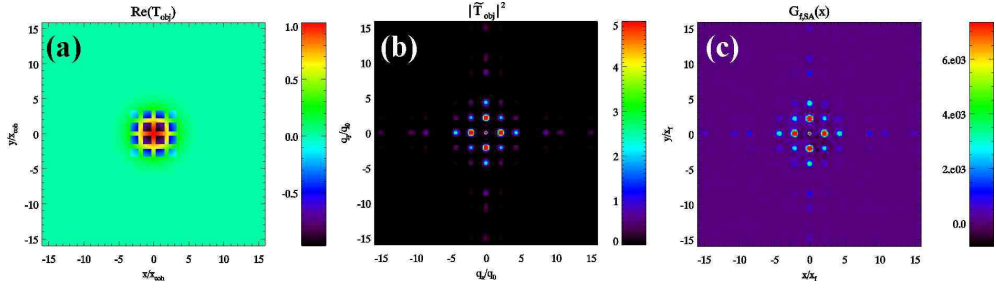}}
%     \scalebox{.29}{\includegraphics{../../OPA/GhostImage/Simulations/2D21a/pattern.ps}}
%     \scalebox{.29}{\includegraphics{../../OPA/GhostImage/Simulations/2D21a/pobj.ps}}
%     \scalebox{.29}{\includegraphics{../../OPA/GhostImage/Simulations/2D21a/final.ps}}
}
\caption{The reconstructed diffraction pattern of a phase object with
  the spatial averaging technique. (a) shows ${\rm Re}(\tobj)$ consisting
  of $4\times 4$ square holes (with the value $\tobj=-1$ inside and
  $\tobj=+1$ outside the holes) modulated by a Gaussian. (b)
  analytical $|\pobj|^2$. (c) $G_{{\rm f},\sa}$ after 2000 shots.}
\label{fig:phase-object}
\end{figure}  

The imaging system with the reference arm set in the f-f setup is able
to reconstruct the diffraction pattern of a pure phase
object,\footnote{Ref.~\cite{abouraddy:2004} shows an experimental
  implementation of imaging the diffraction pattern of a phase object
  in the coincidence counting regime.} both with and without the
spatial averaging technique applied (the latter was already demonstrated
in Ref.~\cite{thermal-oe}). The chosen object [see
Fig.~\ref{fig:phase-object}(a)] was a pure phase object modulated by a
Gaussian as to avoid a cosmetically disturbing DC peak [the analytical
Fourier transform is shown in Fig.~\ref{fig:phase-object}(b)]. An
imaging scheme that is unable to reconstruct phase information would
therefore only show the Fourier transform of the Gaussian which is
another Gaussian.  Figure~\ref{fig:phase-object}(c) shows the result of
a numerical simulation in 2D where the spatial averaging technique was
used, and it closely follows the analytical diffraction pattern of the
object in Fig.~\ref{fig:phase-object}(b) confirming that phase
information is preserved.

\begin{figure}[ht]
\centerline{    
    \scalebox{.55}{\includegraphics{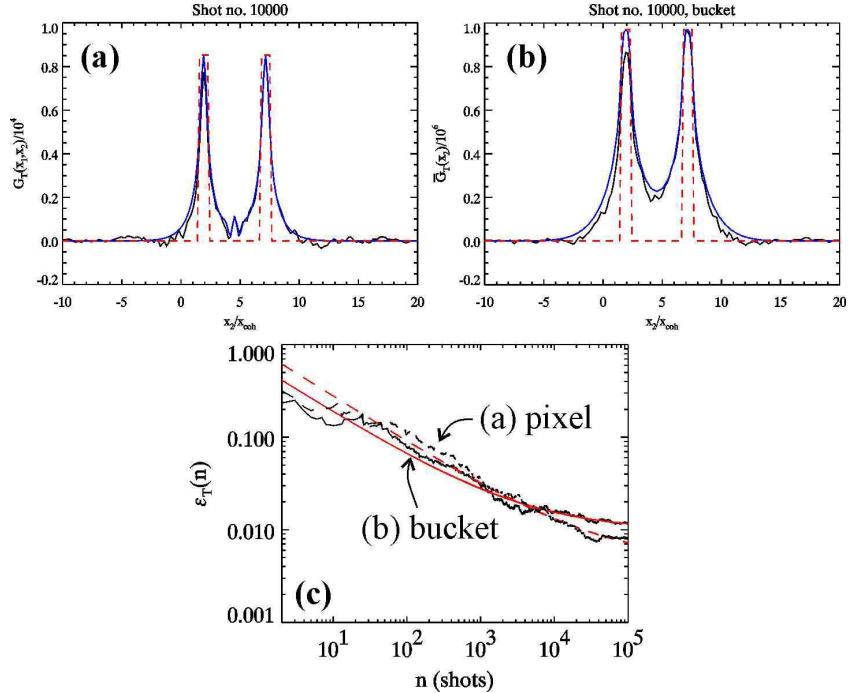}}
%     \scalebox{.27}{\includegraphics{../../OPA/GhostImage/Simulations/1D03/final-nf.ps}}
%     \scalebox{.27}{\includegraphics{../../OPA/GhostImage/Simulations/1D03/final-b.ps}}
%     \scalebox{.5}{\includegraphics{conv1Dnf.eps}}
}
\caption{The reconstructed image using the telescope setup with
  (a) $D_1$ a pixel-like detector and (b) $D_1$ a bucket detector. The
  black curves are numerics, the dashed red curves are $\tobj$, while
  the blue curves are the analytically calculated correlations in the
  SPWPA in arbitrary units [Eqs.~(\ref{eq:telescope-time})
  and~(\ref{eq:telescope-time-bucket})].  (c) shows $\varepsilon_{\rm
    T}(n)$ (full: bucket detector and dashed: pixel detector), and the
  red curves are fits to Eq.~(\ref{eq:exp}). The same parameters and
  notation as in Fig.~\ref{fig:ds}. [Movie sizes: (a) 748 KB and (b)
  1.25 MB].}
\label{fig:ds-nf}
\end{figure}  

The image of the object can be reconstructed by merely changing the
setup in the reference arm to the telescope setup [see
Fig.~\ref{fig:setup}(b)].  Figure~\ref{fig:ds-nf} compares using (a) a
pixel-like detector and (b) a bucket detector in the test arm for this
case. In both cases the double slit is reproduced, albeit with the
characteristic blurring of the sharp contours of the double slit
apertures because of the finite resolution. In fact, the aperture size
of the double slit (14 $\mu$m) is on the limit of the system's
resolution $\xcoh\simeq 16~\mu$m.  The convergence rates as observed
in the animations seem identical, which is confirmed by the error plot
on the right [showing the image version of~(\ref{eq:1}), i.e.
$\varepsilon_{\rm T}(n)= \left(\Sigma_{\x} |G_{\rm {T}}^{n}(\x)-G_{\rm
  {T}}(\x)|^2\right)^{1/2}$]. % This is because the widths of the slits are on the
% order of $\xcoh$, as mentioned above, and hence from
% Eq.~(\ref{eq:conv-nf}) the convergence rate speedup $\rho_{\rm
%   conv}^{\rm NF}$ is roughly unity and thus nothing is gained from
% using a bucket detector. Indeed, by taking the slits 5 times wider the
% bucket detector convergence turned out to be more than 10 times faster
% than using the fixed detector. In addition the wider slits also made
% the convergence slower compared to the thinner slits (with a factor of
% 20 when comparing the fixed detector results of the two different
% objects). This is a confirmation of what was mentioned at the end of
% the previous section, namely that when the object transmission area is
% increased the convergence rate is slower.  

\begin{figure}[ht]
  \centerline{ \scalebox{.55}{\includegraphics{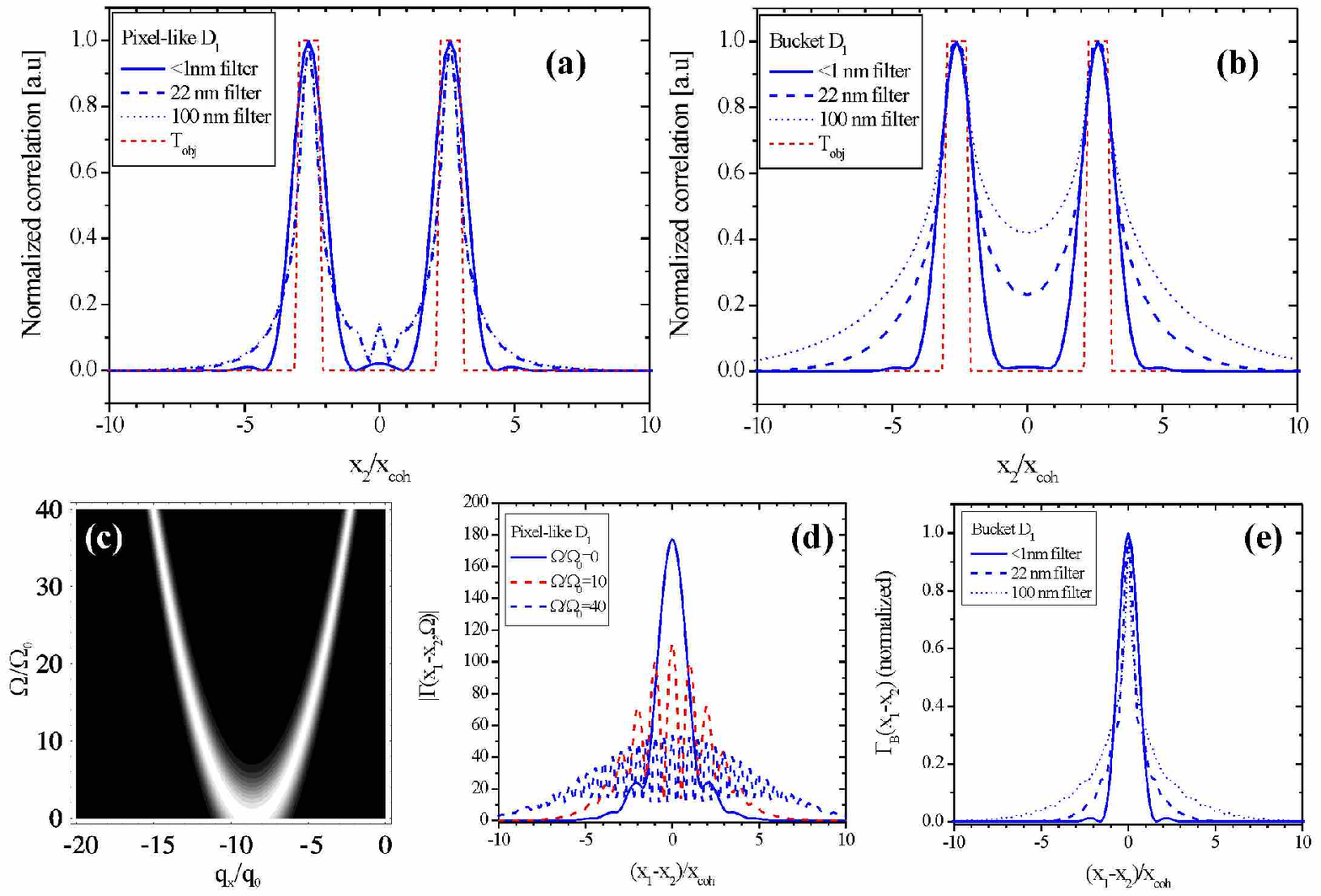}} }
\caption{The reconstructed image from semi-analytical calculations
  with (a) $D_1$ a pixel-like detector and (b) $D_1$ a bucket
  detector, using different interferential filters (different temporal
  bandwidths).  The blue curves are semi-analytical calculations in
  the SPWPA in arbitrary units [Eqs.~(\ref{eq:telescope-time})
  and~(\ref{eq:telescope-time-bucket})]. (c) shows
  $|\gamma(q_x,\Omega)|$ for the chosen PDC setup.  (d) shows
  $|\Gamma(\x_1-\x_2,\Omega)|$ for different values of $\Omega$. (e)
  shows $\Gamma_{\rm B}(\x_1-\x_2)$ for different interferential
  filters.  }
\label{fig:ds-nf-math}
\end{figure}  

From the semi-analytical calculations an interesting observation about
the resolution abilities of the imaging scheme emerged when varying
the interference filter after the PDC crystal, which filters a
frequency range $[-\delta \Omega,\delta \Omega]$. In the case
considered in the analytical section $\delta \Omega=0$, which in
practice corresponds to a $<1$ nm interference filter. The numerics
with time used $\delta \Omega=40 \Omega_0$, i.e. a 22 nm filter.
Figure~\ref{fig:ds-nf-math}(a) shows the case where $D_1$ is a
pixel-like detector; the result remains more or less unchanged as the
interference filter changes and no noticeable difference is observed
between the 22 nm and 100 nm filters. In contrast, when $D_1$ is a
bucket detector the result is much more sensitive; as shown in
Fig.~\ref{fig:ds-nf-math}(b) the narrow interference filter gives the
best resolution, while taking a broader filter gives a deteriorating
resolution. The explanations for these different behaviours are found
from considering the two explicit expressions for the correlations
integrated over time. From Eq.~(\ref{eq:Gtime}) we namely find instead
of Eqs.~(\ref{eq:telescope1}) and~(\ref{eq:telescope1-bucket})
\begin{eqnarray}
G_{\rm T}(\x_1,\x_2)
\propto
\int \dd \Omega\left| \int \dd \xp_1
% \la b_1(\xp_1)b_2(\x_2)\ra
\Gamma(\xp_1-\x_2,\Omega)
 \tobj \left( \xp_1  \right) e^{-\im \xp_1 \cdot \x_1 k/f}
\right|^2, \quad \text{ coherent (pixel-like~$D_1$)}
\label{eq:telescope-time} \\
\bar G_{\rm T}(\x_2)
\propto
\int \dd \Omega \int \dd \x_1
% \la b_1(\xp_1)b_2(\x_2)\ra
|\Gamma(\x_1-\x_2,\Omega)|^2
|\tobj \left( \xp_1  \right) |^2, \quad \text{ incoherent (bucket~$D_1$)}.
\label{eq:telescope-time-bucket}
\end{eqnarray}
Thus, in the coherent case $\Gamma(\xp_1-\x_2,\Omega)$ acts at each
$\Omega$ as an imaging kernel, which is shown in
Fig.~\ref{fig:ds-nf-math}(d) for different values of $\Omega$. As
$\Omega$ increases $\Gamma(\xp_1-\x_2,\Omega)$ becomes broader and the
sidebands become pronounced. This can be understood by noting
that Eq.~(\ref{eq:gammaentangled}) expresses
$\Gamma(\xp_1-\x_2,\Omega)$ as the inverse Fourier transform of
$\gamma(\q,\Omega)$, which is shown in Fig.~\ref{fig:ds-nf-math}(c): as
$\Omega$ increases $\gamma(\q,\Omega)$ becomes double-peaked, giving
the sideband oscillations.  All these contributions should then be
added up as the integration in $\Omega$ is carried out. It is
therefore important to notice that $\Gamma(\xp_1-\x_2,\Omega)$ becomes
weaker as $\Omega$ is increased. Therefore, despite the fact that the
broader sidebands in $\Gamma(\xp_1-\x_2,\Omega)$ imply a deteriorating
resolution, the reduced peak value of $\Gamma(\xp_1-\x_2,\Omega)$
implies that the end result is affected less and less as $\Omega$ is
increased.
% , and more importantly that because the imaging is coherent
% both amplitude and phase of $\Gamma(\xp_1,\x_2,\Omega)$ acts as the
% image kernel in Eq.~(\ref{eq:telescope-time-bucket}). In fact, the
% sideband oscillations of $\Gamma(\xp_1,\x_2,\Omega)$ are both positive
% and negative thus effectively cancelling out the deterioration in
% resolution one should expect from broad sidebands in the image kernel.
Thus, the coherent nature of the imaging explains why the image does
not change much as the bandwidth is increased in
Fig.~\ref{fig:ds-nf-math}(a). In the bucket detector case the situation
is completely different.  Instead of having many contributions from
different $\Omega$, the incoherence of the imaging system implies that
the temporal integration defines a new imaging kernel as
\begin{equation}
  \label{eq:Gamma-bucket}
  \Gamma_{\rm B}(\x_1-\x_2)\equiv\int \frac{\dd
  \Omega}{2\pi}|\Gamma(\x_1-\x_2,\Omega)|^2. 
\end{equation}
so $\bar G_{\rm T}(\x_2) \propto \int \dd \x_1 \Gamma_{\rm
  B}(\x_1-\x_2,\Omega) |\tobj \left( \x_1 \right) |^2$.  Thus,
$\Gamma_{\rm B}(\x_1-\x_2)$ uniquely determines the resolution of the
incoherent imaging system, and from Fig.~\ref{fig:ds-nf-math}(e) we see
it becomes broader as the frequency bandwidth is increased, implying a
decrease in resolution which is exactly what was seen in
Fig.~\ref{fig:ds-nf-math}(b). We should stress that this result is a
particular consequence of the PDC phase-matching conditions and does
therefore not necessarily apply to ghost imaging schemes using other
sources for the correlated beams.

\begin{figure}[ht]
\centerline{    
%%    \scalebox{.35}{\includegraphics{../../OPA/GhostImage/Simulations/2D20c/GTb.ps}}
%    \scalebox{.3}{\includegraphics{../../OPA/GhostImage/Out/GTfix.ps}}
%    \scalebox{.3}{\includegraphics{../../OPA/GhostImage/Out/GTb.ps}}
%     \scalebox{.3}{\includegraphics{../../OPA/GhostImage/Simulations/2D20d/100000/GTfix.ps}}
%    \scalebox{.3}{\includegraphics{../../OPA/GhostImage/Simulations/2D20d/100000/GTb.ps}}
    \scalebox{0.64}{\includegraphics{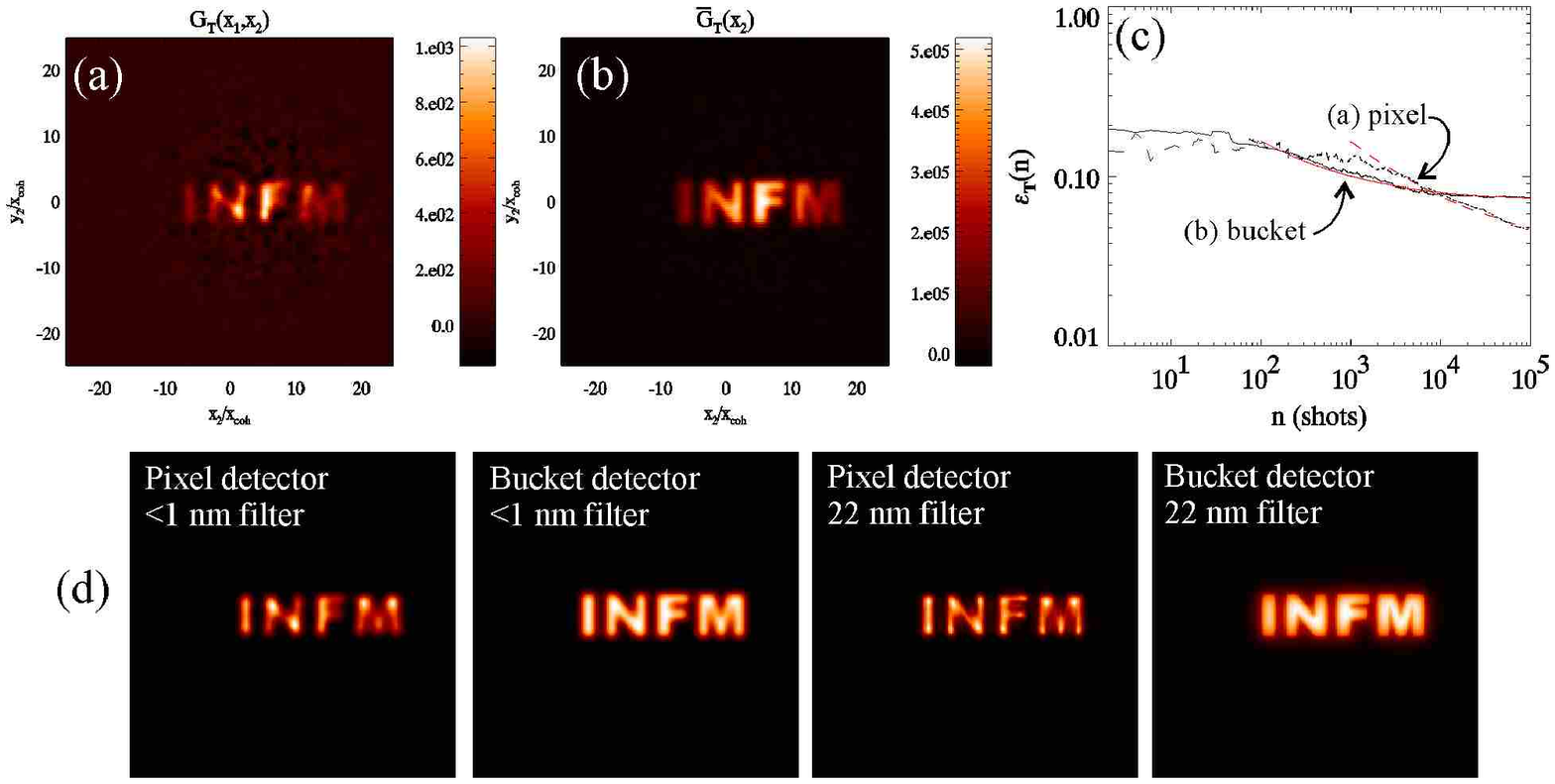}}
}
\caption{The reconstructed near field of a mask with the letters
  ``INFM'' using the telescope setup. In the numerical simulations of
  (a) $D_1$ was a pixel-like detector, while in (b) $D_1$ was a bucket
  detector. (c) shows the convergence rate of the two cases (same
  notation as in Fig.~\ref{fig:ds-nf}.) $10^5$ shots were used, as
  well as a larger pump waist than usual (corresponding to
  $800~\mu$m). (d) shows semi-analytical calculations without time
  ($<1$ nm filter) and with time (22 nm filter).}
\label{fig:infm}
\end{figure}  

In 2D the advantage of using the bucket detector for the image
reconstruction becomes more obvious.
% the near field distribution of relatively complicated objects
% can be reconstructed using the telescope setup, provided that the
% diffraction pattern is not too complicated. By this we mean that the
% diffraction pattern cannot contain significant contributions for large
% $\q$ components, since the near-field imaging system will cut off
% these components. 
As an example of a reasonably complicated object, we have in
Fig.~\ref{fig:infm} reconstructed a mask with the letters ``INFM''.
The difference between the pixel-detector case (a) and the bucket
detector case (b) is again related to coherent vs. incoherent
imaging.\footnote{The curvature of the reproduced images is
  due to the Gaussian profile of the signal field impinging on the
  object.} The semi-analytical calculations in (d) in the SPWPA
confirm this; whereas the pixel-detector result is suffering from a
speckle-like effect \cite{goodman:1968}, this effect is absent in the
bucket-detector result. Note also the different results obtained from
a narrow and a broad bandwidth filter, as discussed in the previous
paragraph [the numerics in Fig.~\ref{fig:infm}(a)-(b) are without
time, so they should be compared to $<1$ nm filter results of
Fig.~\ref{fig:infm}(d)].  Finally, (c) shows $\varepsilon (n)$, and it
is evident that the bucket-detector case (full lines) converges faster
than the pixel-detector case (dashed lines); from the fits we obtain
$d_0^{\rm bucket}/d_0^{\rm pixel}\simeq 20$. This speedup in
convergence is a consequence of the bucket detector $D_1$ making a
spatial average over the field at the test arm detection plane, and
the magnitude of the speedup is therefore related to the number of
independent modes recorded by $D_1$. In 1D this number is much smaller
than in 2D which is why no significant speedup was observed in the 1D
case (Fig.~\ref{fig:ds-nf}).

\begin{figure}[ht]
  \centerline{ \scalebox{.6}{\includegraphics{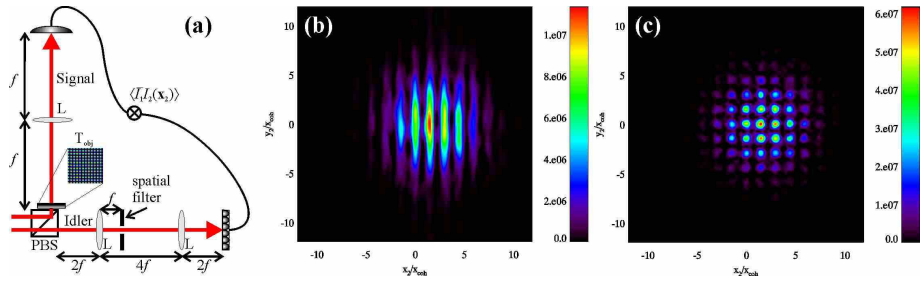}}
% \scalebox{.33}{\includegraphics{filter.eps}}
%     \scalebox{.3}{\includegraphics{ghst2ffBOEa.ps}}
%     \scalebox{.3}{\includegraphics{ghst2fBOEa.ps}} 
}
\caption{Non-local filtering using the telescope setup with a bucket
  detector in the test arm. The setup (a) was as Fig.~\ref{fig:setup}(b)
  but with a spatial filter inserted in the focal plane of the first
  lens in the reference arm. (b) shows the correlation using the
  filter, while (c) shows the correlation without filter. 15000 shots
  were used. %(Movie sizes: 1 MB each)
}
\label{fig:filter}
\end{figure}  

If filters are inserted in the reference arm the reconstructed object
can be manipulated. As a simple example of this, using the telescope
setup in the reference beam we have inserted a filter in the focal
plane of the first lens [see Fig.~\ref{fig:filter}(a)]. The object in
the test arm was a square pattern $T_{\rm
  obj}(\x)=[1+\cos(3xq_0/2)][1+\cos(3yq_0/2)]/4$ (consisting of 4 main
sidebands, 2 located at $\q/q_0=\pm \frac{3}{2}{\bf e}_x$ and 2
located at $\q/q_0=\pm \frac{3}{2}{\bf e}_y$.). Filtering in the
$y$-direction we observe only rolls in the $x$-direction
[Fig.~\ref{fig:filter}(b)], while removing the filter the square pattern
is seen in the correlations [Fig.~\ref{fig:filter}(c)].  This shows that
the image Fourier components can be manipulated non-locally.

\section{Conclusion}
\label{sec:Conclusion}

The ghost imaging schemes rely on two spatially correlated beams.
These are generated by a source with a limited spatial bandwidth in
Fourier space, which determines the highest Fourier component in the
diffraction pattern of the object that can be reproduced (i.e., the
imaging bandwidth is limited). In turn, the resolution of the
reconstructed image is limited by the characteristic near-field
coherence length of the source.

In this paper we have through theory and numerics analyzed in detail a
technique (already presented by us in \cite{bache:2004} for a
different imaging scheme) that improves the imaging bandwidth by
implementing a spatial average over the test arm detector position.
This technique makes the imaging bandwidth of the reconstructed
diffraction pattern virtually infinite (apart from limitations arising
from lenses and apertures), as well as making the correlations
converge much faster. We showed that the speed-up of this convergence
is related to the number of statistically independent modes inside the
source bandwidth. When reconstructing the object image no benefits to
the image resolution could be made by doing a spatial average.
However, by merely using a bucket detector collecting all photons in
the test arm, the imaging becomes incoherent. Whether coherent or
incoherent imaging is advantageous depends obviously on the problem at
hand \cite{goodman:1968}, but we showed examples where the speckle
problem of coherent imaging could be avoided using a bucket detector.
On the other hand, when using a bucket detector it turned out to be
important to use a narrow-band interference filter of the
down-converted beams, otherwise a strong degrading in resolution was
observed. Finally, we also showed that by inserting a filter in the
focal plane of a lens in the reference arm, the reconstructed image
could be non-locally filtered; the filter never interferes with the
field containing the object information but through the correlations
the image is filtered nonetheless.

We have in this paper chosen to focus on an imaging scheme using
spatially entangled PDC beams. However, we should stress that
practically all the results shown are general for any imaging system,
quantum or classical, and are therefore relevant also for the
important case when classical spatially correlated beams are created
by splitting a thermal-like radiation on a beam splitter
\cite{thermal-oe,ferri:2004}. In that case, \textit{e.g.} the
near-field correlation function (\ref{eq:gammaentangled}) is instead
governed by the second-order correlation of the thermal field
\cite{thermal-oe}, but the overall results of this paper remains the
same. Thus, the spatial averaging technique may still be used to
extend the bandwidth and speed up the convergence. In this connection
it should be stressed that with thermal-like beams the reconstructed
object image has exactly the same properties as when using PDC beams:
it is coherent of one uses a pixel-like detector, while it becomes
incoherent if a bucket detector is used.

\section*{Acknowledgments}
\label{sec:Acknowledgments}

This project has been carried out in the framework of the FET project
QUANTIM of the EU, of the PRIN project MIUR ``Theoretical study of
novel devices based on quantum entanglement'', and of the INTAS
project ``Non-classical light in quantum imaging and continuous
variable quantum channels''. M.B. acknowledges financial support from
the Danish Technical Research Council (STVF) and the Carlsberg Foundation.

%\begin{figure}
%\centerline{\includegraphics{OpEx2.EPS}}
%\caption{}
%\end{figure}

\end{document}